\documentclass[useAMS,usenatbib]{mn2e}
\def \lcdm {$\Lambda$CDM}
\def \lwdm {$\Lambda$WDM}

\pdfoutput=1
\usepackage{amsfonts,amssymb,amsmath}
\usepackage{aas_macros}
\usepackage{times,varioref,multirow,textcomp,comment}
\usepackage[usenames,dvipsnames,svgnames,hyperref]{xcolor}
\usepackage{comment}
\usepackage[pdftex]{graphicx}
\usepackage[
    pdftex,
    a4paper=true,
    plainpages=true,
    pdfpagelabels,
    breaklinks=true,
    bookmarks=true,
    bookmarksopen=false,
    bookmarksopenlevel=2
    bookmarksnumbered=true,
    bookmarkstype=toc,
    colorlinks=true,
    citecolor=RoyalBlue,
    linkcolor=ForestGreen,
    menucolor=Teal,
    urlcolor=DarkOrange,
]{hyperref}

\providecommand{\adsurl}[1]{\href{#1}{ADS}}

\hypersetup{
  pdfauthor = {Hareth Mahdi, Pascal Elahi, Chris Power, Geraint Lewis, Martijn van Beek, Madhura Killedar},
  pdfkeywords = {Lensing, Warm Dark Matter},
  pdftitle = {}
}

\def \Msun{\ {\rm M_\odot}}
\def \Msunh{\ h^{-1}{\rm M_\odot}}

\title[Gravitational lensing in WDM cosmologies]{Gravitational lensing in WDM cosmologies: The cross section for giant arcs }
\author[Hareth S. Mahdi et al.]
{\parbox{\textwidth}{Hareth S. Mahdi$^{1,2}$\thanks{E-mail: hareth@physics.usyd.edu.au},
Martijn van Beek$^{1,3}$,
Pascal J. Elahi$^{1}$, 
Geraint F. Lewis$^{1}$,
Chris Power$^{4}$,  
and Madhura Killedar$^{5}$}\vspace{0.4cm}\\
\parbox{\textwidth}{
$^{1}$Sydney Institute for Astronomy, School of Physics, A28, The University of Sydney, NSW 2006, Australia\\
$^{2}$Department of Astronomy, University of Baghdad, Jadiryah, Baghdad 10071, Iraq \\
$^{3}$Institute for Mathematics, Astrophysics and Particle Physics, Radboud University Nijmegen, Heyendaalseweg 135, 6525 GL Nijmegen, The Netherlands\\
$^{4}$International Centre for Radio Astronomy Research, University of Western Australia, 35 Stirling Highway, Crawley, WA 6009, Australia\\
$^{5}$Universit\"ats-Sternwarte M\"unchen, Scheinerstrasse 1, D-81679, M\"unchen, Germany\\
}
}

\begin{document}

\date{}

\pagerange{\pageref{firstpage}--\pageref{lastpage}} \pubyear{2013}

\maketitle

\label{firstpage}

\begin{abstract}
The nature of the dark sector of the Universe remains one of the outstanding problems in modern cosmology, with the search for new observational probes guiding the development of the next generation of observational facilities. Clues come from tension between the predictions from \lcdm\ and observations of gravitationally lensed galaxies. Previous studies showed that galaxy clusters in the \lcdm\ are not strong enough to reproduce the observed number of lensed arcs.   
This work aims to constrain the warm dark matter cosmologies by means of the lensing efficiency of galaxy clusters drawn from these alternative models. The lensing characteristics of two samples of simulated clusters in the warm dark matter (\lwdm) and cold dark matter (\lcdm) cosmologies have been studied. The results show that even though the CDM clusters are more centrally concentrated and contain more substructures, the WDM clusters have slightly higher lensing efficiency than their CDM counterparts. The key difference is that WDM clusters have more extended and more massive subhaloes than CDM analogues. These massive substructures significantly stretch the critical lines and caustics and hence they boost the lensing efficiency of the host halo. Despite the increase in the lensing efficiency due to the contribution of massive substructures in the WDM clusters, this is not enough to resolve the arc statistics problem.      
\end{abstract}


\section{Introduction}
\label{sec:intro}
The standard model of cosmology (Cold Dark Matter + Dark Energy) is found to be in good agreement with the observational data on large scales of the Universe \cite[e.g.][]{2005MNRAS.362..505C,2010MNRAS.401.2148P,2011ApJS..192...18K,2013MNRAS.436.1674A}. However, there are a few discrepancies between observations and the CDM model. For instance, the \lcdm\ cosmology predicts that many more satellites must be around Milky Way-sized galaxies than the observed satellites \citep{1999ApJ...522...82K,1999ApJ...524L..19M}. This problem could be resolved by adopting warm dark matter model as an alternative to the standard model \cite[e.g.][]{2000ApJ...542..622C,2001ApJ...556...93B,2014MNRAS.tmp..192L}.

Gravitational lensing has revealed another discrepancy between observation and the current cosmological model. Galaxy clusters are observed to produce more lensed giant arcs than predicted by the \lcdm\ model \cite[for a review see][]{2013SSRv..177...31M}. This discrepancy is known as the arc statistics problem and was firstly addressed by \cite{1998A&A...330....1B}. 
Further work from \cite{2004ApJ...606L..93W} was showed that the lensing optical depth for giant arcs is strongly dependent on the source redshift due to the very steep cluster mass function. Contrary to \cite{1998A&A...330....1B}, these authors have found that the abundance of giant arcs in the \lcdm\ model agrees with that of the observed clusters when using a wider range of source redshift. 
However, later research by \cite{2005ApJ...635..795L} confirmed the arc statistics problem and found that the optical depth increased at a slower rate with the source redshift than reported in \cite{2004ApJ...606L..93W}.  
A recent study from \cite{2011A&A...530A..17M} has confirmed the findings of \cite{1998A&A...330....1B}. However the new study shows that the arc statistics discrepancy between the lensing efficiency of the observed clusters and the \lcdm\ clusters is about a factor of two rather than one order of magnitude. We note that the lensing efficiency can be significantly boosted during the cluster mergers, where the lensing cross section can increase by one order of magnitude or more on a time scale of $\sim$ 0.1 Gyr \citep{2004MNRAS.349..476T}. It has also been found that the \lcdm\ clusters produce smaller Einstein radii than those observed \cite[e.g.][]{2008MNRAS.390.1647B, 2011MNRAS.410.1939Z}. The abundance of giant arcs is found to be sensitively dependent on the concentration of matter in their cores, and hence provides a statistical constraint on the density profile of galaxy clusters \cite[e.g.][]{1993MNRAS.262..187W,2001ApJ...559..572O}. The optical depth for giant arcs could be boosted drastically by increasing the normalisation of power spectrum \citep{2008A&A...486...35F}. 
\cite{2009MNRAS.398.1298P} investigated the contribution of structures along the line of sight to the strong lensing efficiency. They found that the abundance of giant arcs for individual clusters could be increased by up to $\sim 50$ per cent, showing that its contribution becomes more significant for clusters of lower masses and sources at higher redshifts. 
Baryonic processes can steepen the central density profile of galaxy clusters and hence increase lensing efficiency by a small factor. However, that is not enough to resolve the dicrepancy observed \cite[e.g.][]{2005A&A...442..405P,2008ApJ...687...22R,2008ApJ...676..753W,2010MNRAS.406..434M,2012MNRAS.427..533K}. 

These studies have all used CDM, however, the fundamental nature of dark matter is still a subject of debate. The dark matter in the Universe must be non-baryonic, but the important question that needs to be answered is: Is the dark matter hot, cold or warm? 
Hot dark matter scenario has been ruled out in the early 1980s as the free streaming scale is found to be too large and hence galaxies would not form \citep{1983ApJ...274L...1W}. On the other hand, the structure formation in the Universe based on the cold dark matter scenario has been a great success at describing the large scales of the Universe (e.g. \citealt{1984Natur.311..517B, 1985ApJ...292..371D}). The \lwdm\ cosmology has started to receive more attention in the last several years as it agrees better with observations on small scales \cite[e.g.][]{2012MNRAS.420.2318L,2012MNRAS.424..684S,2013PASA...30...39L}. The differences between the structure formation of the CDM and WDM have been outlined by \cite{2001ApJ...556...93B}. These authors have pointed out that the WDM structures have larger core radii, lower core densities and less abundance of substructures. 

Therefore, it is only natural to study whether WDM is a solution to the arc statistics problem. In the present work, we explore both \lcdm\ and \lwdm\ cosmologies to improve our understanding about the differences of structures between the two cosmologies and how those differences affect the lensing efficiency to produce giant arcs. Naively, one would expect that the WDM clusters produce smaller Einstein radii and  cross sections than their CDM counterparts as they are less concentrated and contain fewer substructures. This should result in smaller convergence and shear fields which in turn lead to a lower lensing efficiency. However, our results show that the WDM clusters are slightly more efficient than the CDM analogues due to some physical differences between these two cosmologies. To confirm these physical differences, we compare in a companion study \cite{Elahi2014} the mass accretion and internal properties of the WDM and CDM clusters. 

Our paper is organised as follows: In Section 2, we describe the numerical methods used in this work. We then present the main results of this study (Section 3) and a brief description on why the WDM clusters are slightly stronger lenses than the CDM counterparts (Section 4). Finally, we discuss our results and conclude in Section 5.

\section{Numerical Methods}
\label{sec:num_meth}
\subsection{Simulations}
We study 10 pairs of clusters extracted from zoom simulations (see Table~\ref{table:clusters_properties} for their bulk properties at $z$=0). These zoom simulations used a parent simulation of  $L_{\rm box}=150 h^{-1} \rm Mpc$ containing $128^3$ particles in the \lcdm\ model ($h=0.7$, $\Omega_m=0.3$, $\Omega_{\Lambda}= 0.7$, and $\sigma_8=0.9$). Clusters with masses of $>10^{14}\Msunh$ were identified in the parent simulation using {\small AHF}, which uses a overdensity threshold approach to identify halos ({\small{\textbf{A}MIGA}'s 
  \small{\textbf{H}}alo \small{\textbf{F}}inder}; cf. \citealt{2009ApJS..182..608K}). The overdensity used to define the halo corresponded to the so-called virial radius $r_{\rm vir}$, where the mean interior density is $\Delta_{\rm vir}$ times the critical density of the Universe at that redshift, $\rho_{\rm c}(z)=3H^2(z)/8\pi G$, where $H(z)$ and $G$ are the Hubble parameter and the gravitational constant, respectively. For ten such clusters, we identified all particles within a radius of $\sim 3 R_{\rm vir}$ of the cluster at $z$=0 in the parent simulation and determined their initial positions using a inverse Zel'dovich transformation to obtain the particle positions at $z=\infty$, from which we determined the spatial extent of the initial Lagrangian volume. This volume defines the central region of a multi-level mask for the high resolution region.
The resampled lagrangian regions are initialised using \lcdm\ and \lwdm\ cosmologies, which primarily differ in the power included at small scales. We present here a brief description of the \lwdm\ model, for more details see \cite{Elahi2014}. Our \lwdm\ model  used is a 0.5 keV thermally produced dark matter particle \citep{2001ApJ...556...93B}, which results in a suppression of growth for halo with $M \lesssim M_{hm} = 2.1 \times 10^{11} \Msun$, the so-called half-mode mass scale where the WDM power spectrum is 1/4 that of the CDM one \citep{2012MNRAS.424..684S}. All simulations are run with {\small GADGET2}, a TreePM code \citep{2005MNRAS.364.1105S} and the zoom simulations used a gravitational softening length based on \cite{2003MNRAS.338...14P}, {\it i.e.}, $\epsilon_{\rm opt}=4\,r_{\rm vir}/\sqrt{N_{\rm vir}}$.

\begin{table}
\centering
    \caption{Cluster Properties at $z$=0. $M_{\rm vir}$ is the 
      virial mass of the halo, expressed in units of 10$^{15}$ h$^{-1} \rm \Msun$, 
   	   assuming $\Delta_{ vir}$=200; $R_{ vir}$ is the virial radius, in units of 
     h$^{-1}$  Mpc; and $N_{DM}$ is the number of dark matter particles within 
      	$R_{ vir}$.}
    \vspace*{0.3 cm}
    
    \begin{tabular}{lccc}
      \hline
      & $M_{ vir}$ & $R_{ vir}$ & $N_{ DM}$ \\
      & {\scriptsize $10^{15}$ h$^{-1} \Msun$}  & {\scriptsize h$^{-1}$  Mpc} & \\
      
      \hline
    &          &       &        \\
    &          &       &        \\
C1  &  1.02357 & 2.06  & 681684 \\ 
C2  &  0.74462 & 1.85  & 507416 \\
C3  &  0.62294 & 1.74  & 423850 \\
C4  &  0.52151 & 1.64  & 357024 \\
C5  &  0.49527 & 1.61  & 322581 \\
C6  &  0.43871 & 1.55  & 309802 \\
C7  &  0.42237 & 1.53  & 298265 \\
C8  &  0.40659 & 1.51  & 297888 \\
C9  &  0.40326 & 1.51  & 269994 \\
C10 &  0.41351 & 1.52  & 292005 \\

\label{table:clusters_properties}
\end{tabular}
\end{table}

\subsection{Ray tracing method} 
\label{sec:ray_trace}
Mock lensing maps for two samples of simulated clusters in WDM and CDM cosmologies were produced using the ray tracing method. Mapping the light rays from the lens plane to the source plane can be done using the lensing equation:
\begin{equation} \label{eq:lensequation}
    {\bf \beta}={\bf \theta}-{\bf \alpha(\bf \theta)},
\end{equation} 
where $\bf \beta$ and $\bf \theta$ are the angular positions of light rays on the source and lens planes, respectively and $\bf \alpha(\bf \theta)$ is the deflection angle of light rays on the lens plane, which is given by:
\begin{equation}\label{eq:defl_angl}
    {\bf \alpha(\bf \theta)} =\frac{1}{\pi}\int  \kappa({\bf \theta'}) \frac{\bf \theta-\theta'}{{\bf |\theta-\theta'|}^2} \,\mathrm{d}^{2} \theta^{'},
\end{equation}
where $\kappa$ is the convergence. For gravitational lens under the thin lens approximation, the convergence is simply proportional to the projected surface mass density, $\kappa = \frac{\Sigma}{\Sigma_{crit}}$, where $\Sigma_{crit}$ is the critical surface mass density that depends on the angular diameter distances between observer and lens $D_L$, between observer and source $D_S$ and between lens and source $D_{LS}$: 
\begin{equation}\label{eq:crit_sigma}
\Sigma_{crit} = \frac{c^2}{4 \pi G} \frac{D_S}{D_L D_{LS}},
\end{equation}
Note that the thin lens approximation hold as clusters, the gravitational lenses studied here, are much smaller than the above-mentioned distances. For this study, we place our lens and source planes at redshifts of $z_L=0.3$ and $z_S=2.0$ respectively. 

\par
The deflection angle is evaluated by multiplying the convergence map with the kernel from equation~\ref{eq:defl_angl} in Fourier space, that is 
\begin{equation}\label{eq:convolution}
    \bf \alpha(\bf \theta) = \frac{1}{\pi} \left[\kappa(\bf \theta) * K(\bf \theta)\right],
\end{equation}
where
\begin{equation}\label{eq:kernel}
    \bf K(\bf \theta) = \frac{\bf \theta}{|\bf \theta|^2}.
\end{equation}

We implement a zero-padding method to overcome the periodicity of Fourier transform when convolving the convergence with the kernel function. 

\par
For our clusters, we use only particles within $=2.5$Mpc of the cluster centre to calculate the convergence by projecting them onto two 2D grids: The particles within $0.5$Mpc are projected onto a small high resolution grid of (1024 x 1024) such that the angular resolution is 0.22 arcsec and the rest of particles are projected onto a larger lower resolution grid of $5$Mpc of 1024 x 1024.  The projected surface mass density is smoothed using a truncated Gaussian kernel of size of 5 h$^{-1}$kpc in order to overcome the numerical noise due to the discreteness of N-body simulation. 
The contribution of particles in the outer grid to the deflection angle and shear fields of the inner grid has been taken into consideration by implementing a bilinear interpolation scheme between both maps. This technique was also used in \cite{2012MNRAS.427..533K} and its advantage is that it actually produces high resolution lensing maps for the core of clusters where the strong lensing regime can happen. 

\par 
The magnification of an image is characterized by the Jacobian matrix that can be writen in terms of the convergence and the two components of shear $\gamma_1$ and $\gamma_2$ as follows:
\begin{align}
  \mathcal{A} = \frac{\partial\boldsymbol{\beta}}{\partial\boldsymbol{\theta}}
      &=\left(
      \begin{array}{ccc}
          1 - \frac{\partial^2 \psi}{\partial \theta_x^2 } & \frac{\partial^2 \psi}{\partial \theta_x \partial \theta_y} \\
          \frac{\partial^2 \psi}{\partial \theta_x \partial \theta_y} & 1 - \frac{\partial^2 \psi}{\partial \theta_y^2 }
      \end{array}
      \right)\notag\\
      &=\left( 
      \begin{array}{ccc} 
          1- \kappa - \gamma_1 & - \gamma_2 \\
          - \gamma_2 & 1 - \kappa + \gamma_1  
      \end{array}
      \right)
\end{align}
where $\psi$ is the lensing potential: 
\begin{equation}\label{eq:potential}
    \psi=\frac{1}{\pi} \int \kappa({\bf \theta^{'}}) \ln |{\bf \theta}-{\bf \theta^{'}}| \,\mathrm{d}^{2} \theta^{'}.
\end{equation}
The total magnification of an image is given by $\mu = \frac{1}{det|\mathcal{A}|}$. Regions of high magnification on the lens plane occur whenever the determinant of Jacobian matrix vanishes. The two curves associated with this criteria are the radial critical curve where the radial eigenvalue $\lambda_r=1 - \kappa + \gamma$ goes to zero and images are radially elongated with respect to the curve and the tangential critical curve (a.k.a Einstein curve) where the tangential eigenvalue $\lambda_t=1 - \kappa - \gamma$ vanishes and images are tangentially elongated with respect to the curve. The corresponding curves to the critical lines on the source plane are the radial and tangential caustics. 

\par
Associated to the tangential critical curve is the Einstein radius of a lens. This radius is defined as the size of the tangential critical line, even though it happens to be a circle only for axisymmetric smooth mass distribution. The literature contains several techniques to estimate Einstein radius. For instance, \cite{2008MNRAS.390.1647B} define Einstein radius as the projected radius within which the mean convergence equal to one ($\bar{\kappa}$=1), \cite{2011A&A...530A..17M} on the other hand define Einstein radius as the median distance of critical points on the tangential critical line with respect to the center of lens. Other studies use $\sqrt{\frac{A}{\pi}}$ to measure Einstein radius, where $A$ is the area enclosed by the tangential critical line (e.g. \citealt{2009MNRAS.398.1298P}). \cite{2012A&A...547A..66R} refer to the last two definitions as median and effective Einstein radius, respectively. These authors pointed out that the agreement between these two definitions is moderate with systemically smaller effective Einstein radius than the median one. This proves that the median Einstein radius captures the significant effect of lens ellipticity in comparison to the effective Einstein radius. Furthermore, the effect of a more pronounced shear field from substructures is to push the tangential critical points outward where the convergence is comparably low, leading to more elongated Einstein radii. For these two reasons, we decided to estimate Einstein radius using the definition used in \cite{2011A&A...530A..17M}. 

As we have already mentioned, subhaloes can stretch Einstein radius if they are relatively close to the cluster's core. However, if those strong substructures are relatively faraway from the core of the clusters, they can develop their own critical lines. We estimate Einstein radius as the size of tangential critical curve that corresponds to the cluster's core, where the density of a cluster is high enough for the strong lensing regime to be occurred. 

\par
To account for the limited number of clusters we have, we compute the lensing cross section for 150 different lines-of-sight (los). Since individual clusters will have different central concentrations and triaxiality \cite[e.g.][]{2004ApJ...609...50D,2005ApJ...632..841O}, using a single los would bias our results significantly.

\par
To study the strongly lensing characteristics of our clusters we calculate the cross section for giant arcs, defined as the area on the source plane where a source must be located in order to be lensed as a giant arc. For each los, we throw down thousands of sources close to the caustics, where we expect the sources to be lensed as giant arcs. Those sources are assumed to follow a uniform ellipticity distribution from [0:0.5] with equivalent radius of size 0.5 arcsec and random orientations. We then compute the lensing cross section as the area on the source plane covered by sources that are lensed as giant arcs to the image plane. 

\section{Results}
\label{sec:resluts}
\subsection{Mock lensing maps}
\begin{figure*}
\centering  
\includegraphics[width=2.0\columnwidth , height=2.5\columnwidth]{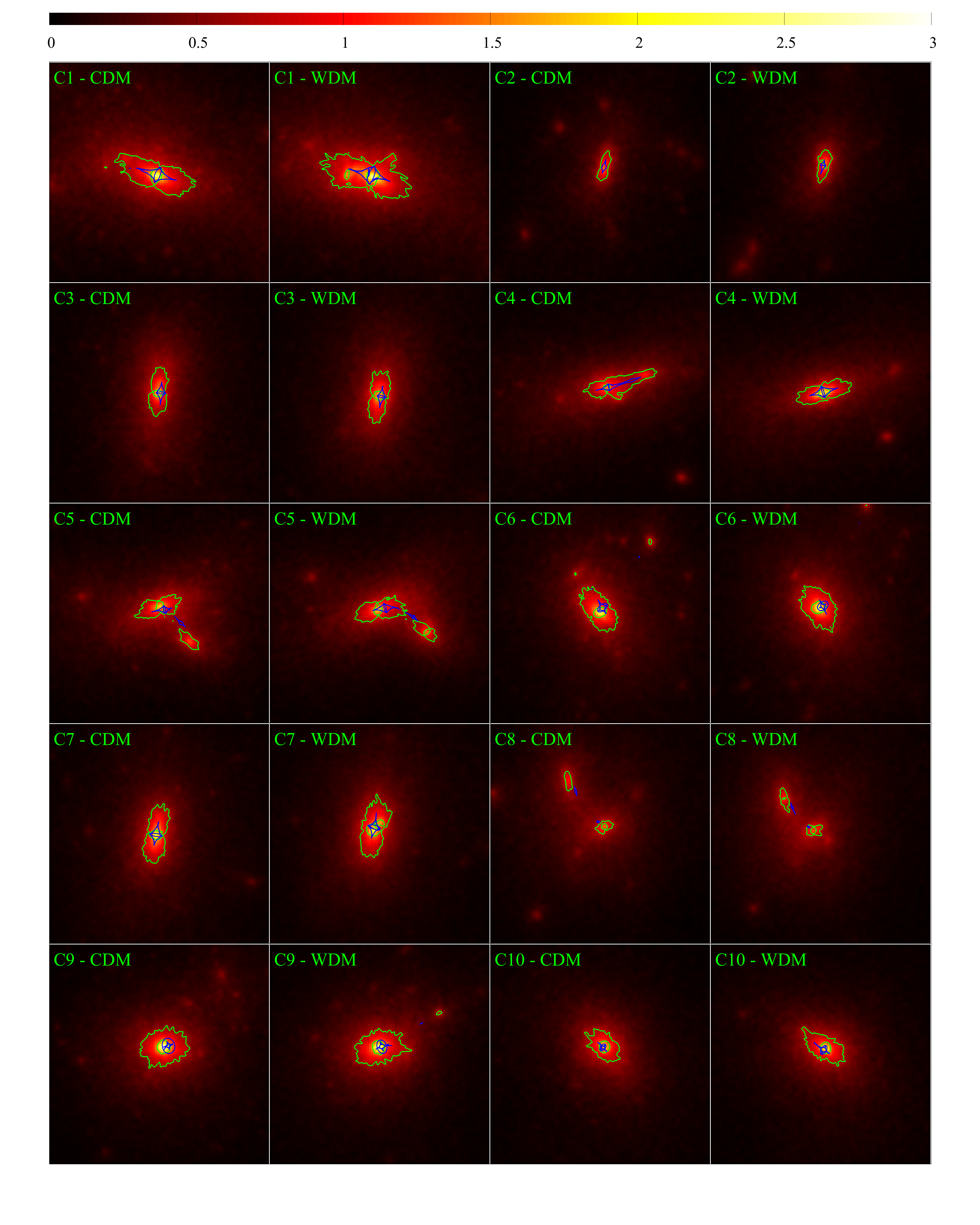}
\caption{The convergence maps for CDM and WDM clusters by considering the rotation matrix that produces the median value of cross section for WDM clusters, the green lines are the critical lines on the lens plane and the blue lines represent the corresponding caustics on the source plane. The side length of each grid is 1 Mpc and the color bar has been set to be in the range [0:3] for all panels. }
\label{fig:med_sig_all}
\end{figure*}

Figure~\ref{fig:med_sig_all} shows the convergence maps for all clusters. Here the green and blue lines are the critical lines on the lens plane and the caustics on the source plane, respectively. This figure illustrates the effect of the larger substructures in the WDM version of some clusters on the critical lines and caustics. Note that here we have chosen to show the orientation that corresponds to the median value of cross section for WDM clusters in the following figures. Similar maps can be done for the lensing shear and deflection angle.  
\begin{figure}
  \centering
  \includegraphics[width=0.99\columnwidth]{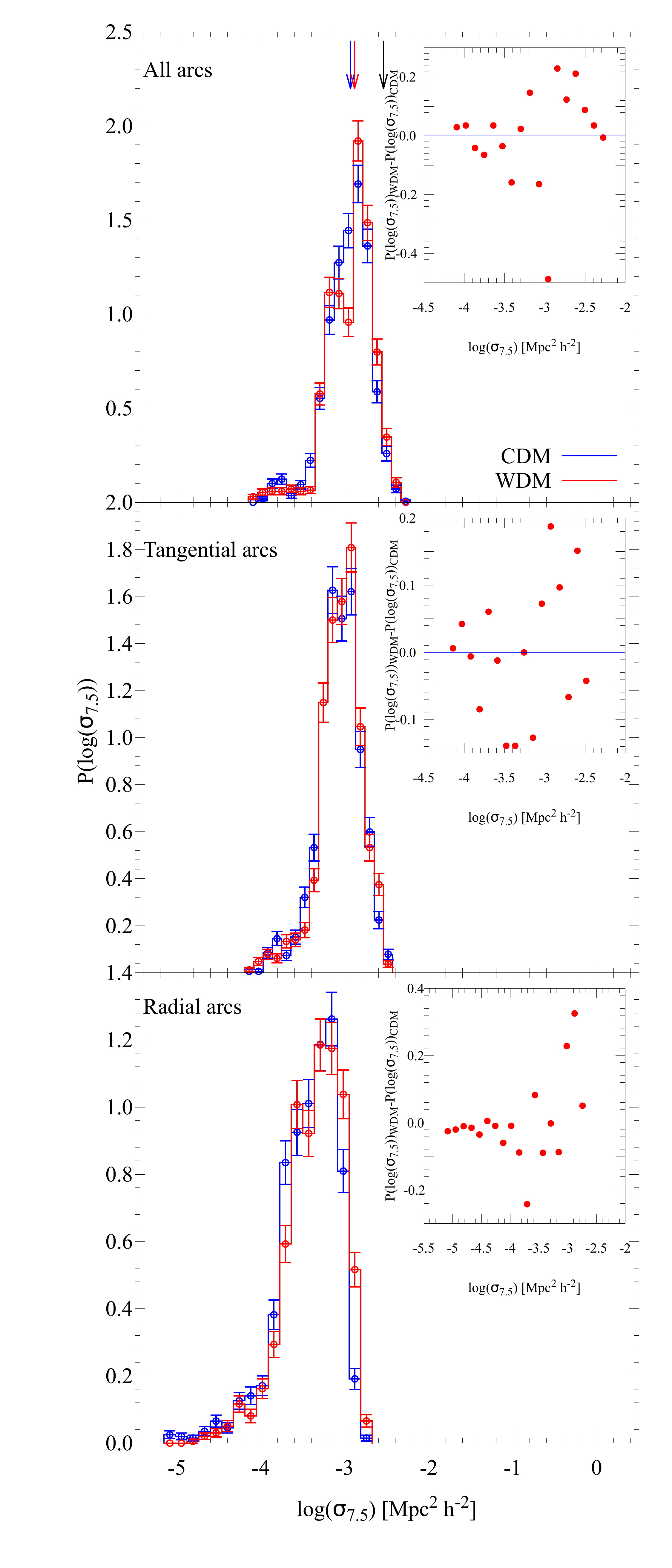}
  \caption{The probability distribution function of cross sections due to all giant arcs (upper panel), tangential arcs only (middle panel), and the radial arcs only (lower panel). The PDF of CDM clusters is shown in (blue) and the counterpart PDF of WDM clusters is shown in (red), the blue, red and black arrows point to the median cross section of the CDM and WDM samples as well as the observed sample in \citep{2011A&A...530A..17M}. The error bars represent the statistical error for each bin and the sub-plots show the difference between the two distributions.}
\label{fig:sigma_PDF}
\end{figure}

\begin{figure}
  \centering  
  \includegraphics[width=0.99\columnwidth]{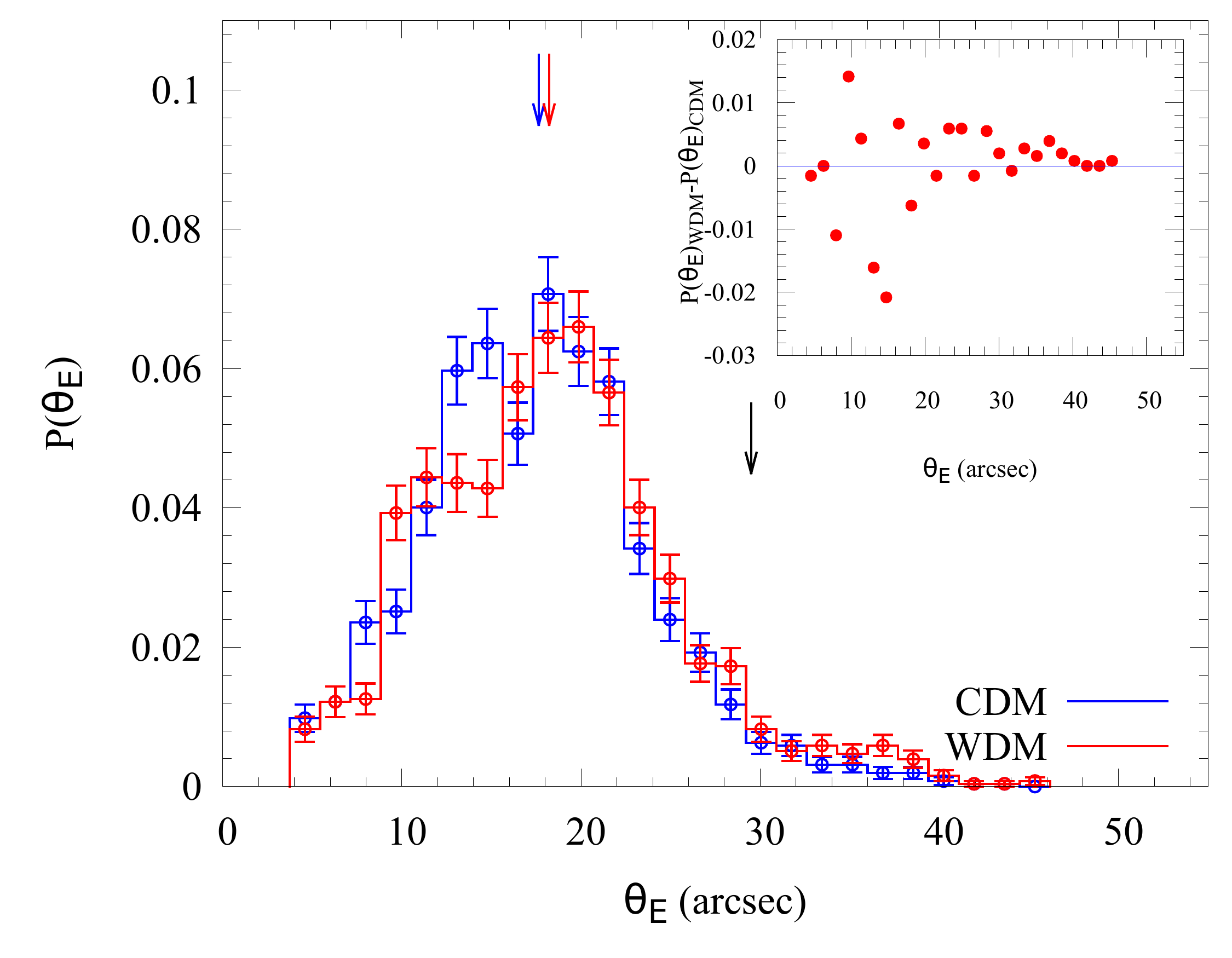}
  \caption{The probability density function of Einstein radii for CDM sample (blue) and WDM sample (red). The blue, red and black arrows point to the median Einstein radius of the CDM and WDM samples as well as the observed sample in \citep{2011A&A...530A..17M}. The error bars represent the statistical error for each bin and the sub-plot shows the difference between the two distributions.}
\label{fig:theta_PDF}  
\end{figure}

\begin{figure}
  \centering  
  \includegraphics[width=0.99\columnwidth]{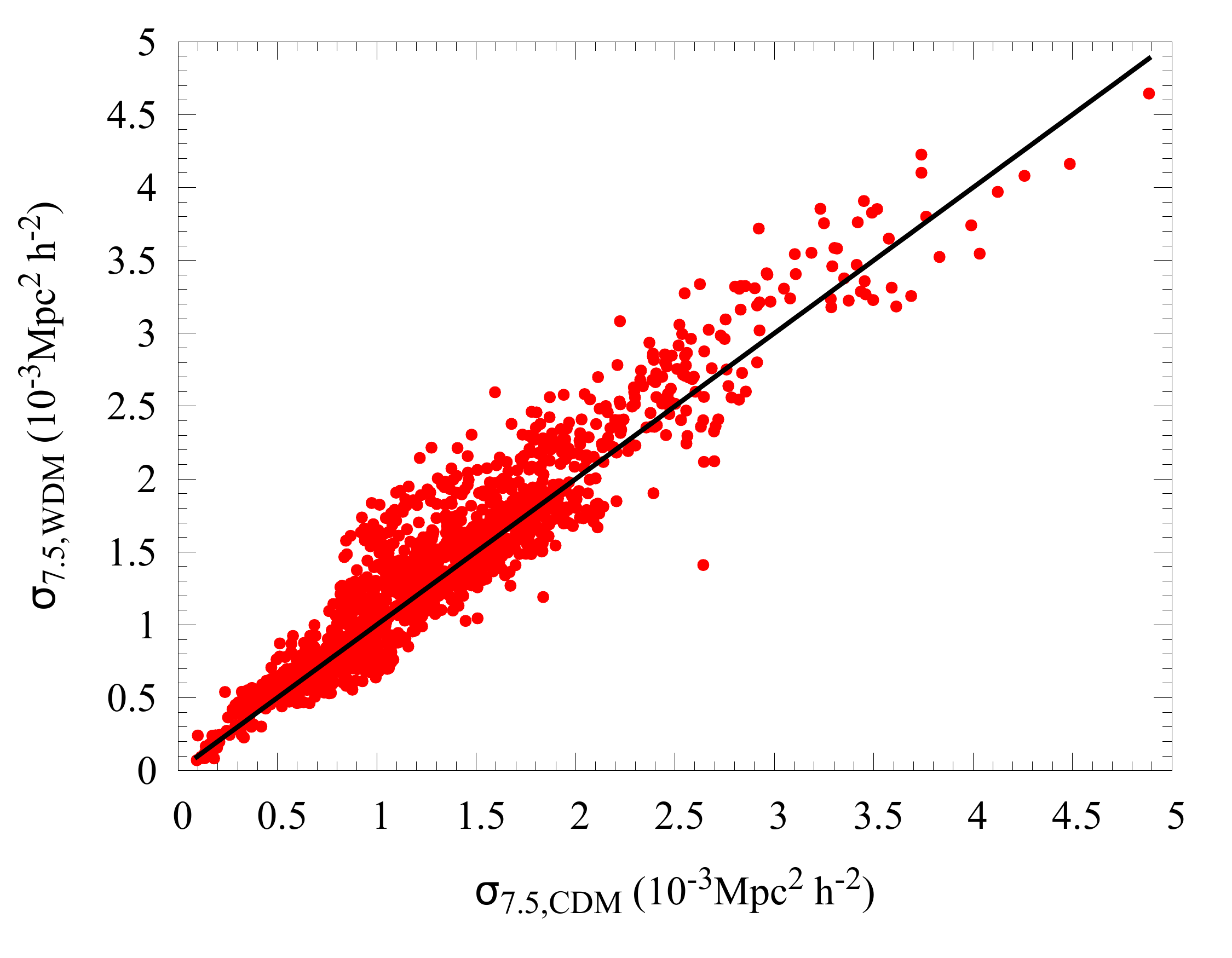}
  \caption{The lensing cross sections of the 1500 los for the WDM version of all clusters versus the cross sections of the CDM counterparts, the black line is the unity line.}
\label{fig:coldvswarm_sigma}  
\end{figure}

\begin{figure}
  \centering  
  \includegraphics[width=0.99\columnwidth]{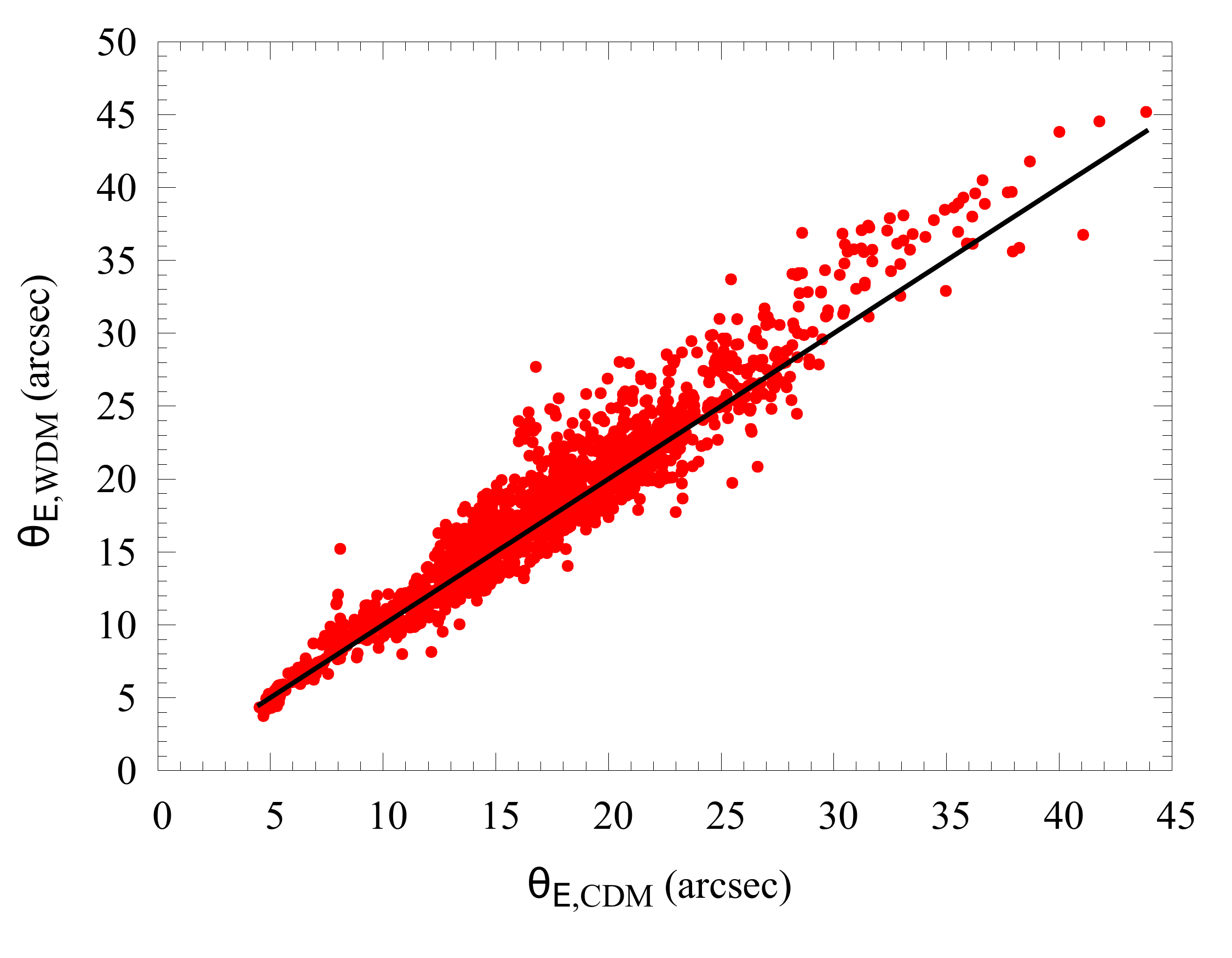}
  \caption{Einstein radii of the 1500 los for the WDM version of all clusters versus Einstein radii of the CDM counterparts, the black line is the unity line.}
\label{fig:coldvswarm_theta}  
\end{figure}

\begin{figure}
  \centering 
  \includegraphics[width=0.99\columnwidth]{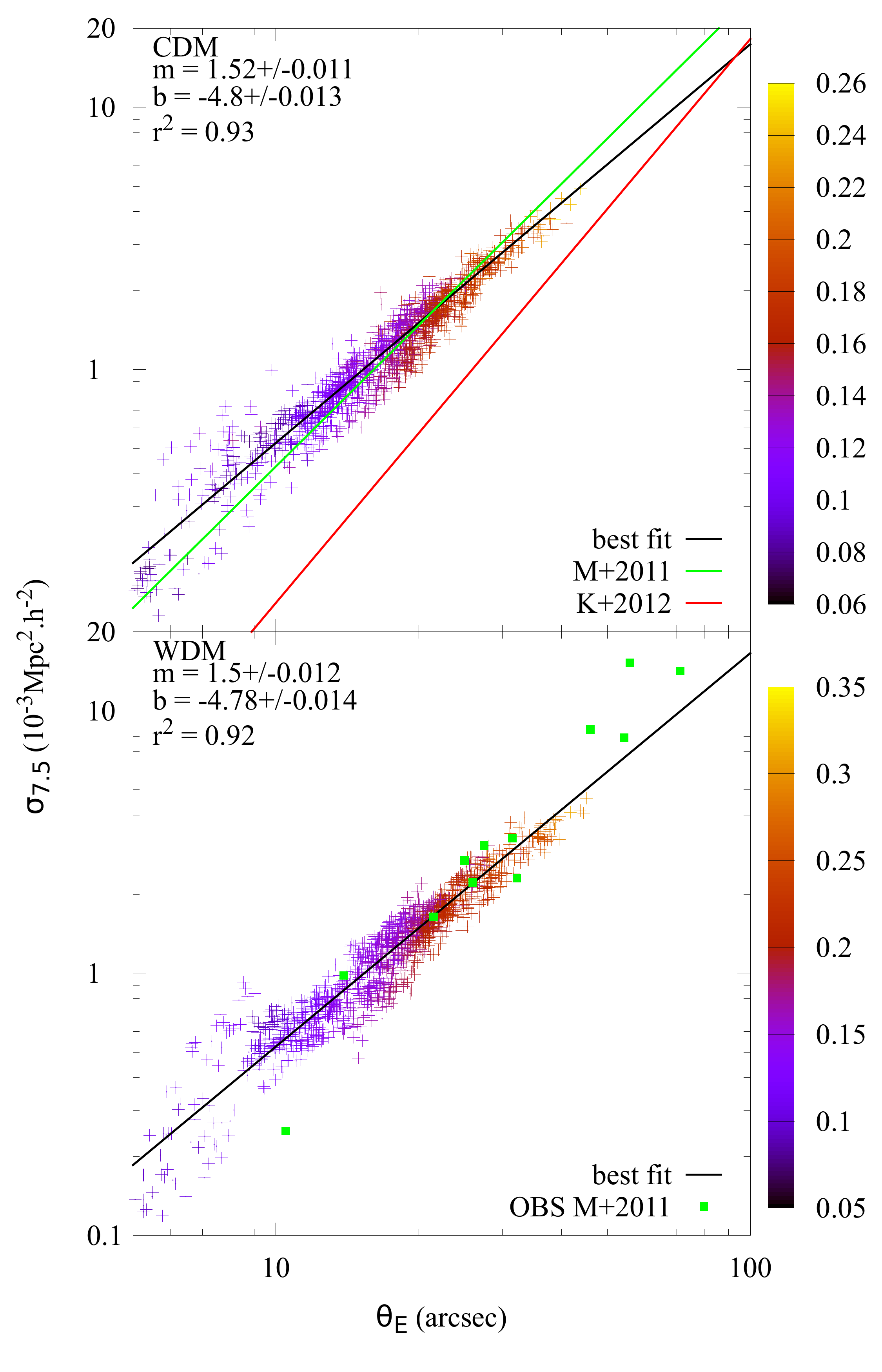}
  \caption{The correlation between the lensing cross sections and Einstein radii for CDM clusters (upper panel) and WDM clusters (lower panel). The black line in both panels shows the best linear fitting to the data. The green line in the upper panel shows the best linear fitting of \citet{2011A&A...530A..17M} and the red and blue lines show the best linear fit of \citet{2012MNRAS.427..533K} by considering the subsample of relaxed clusters and the whole sample, respectively. The data points are coloured with the corresponding mean shear on the grid. The green squares in the lower panel are the observational data from \citet{2011A&A...530A..17M}}.
\label{fig:fitting}
\end{figure}
\begin{figure}  
  \centering  
  \includegraphics[width=0.99\columnwidth]{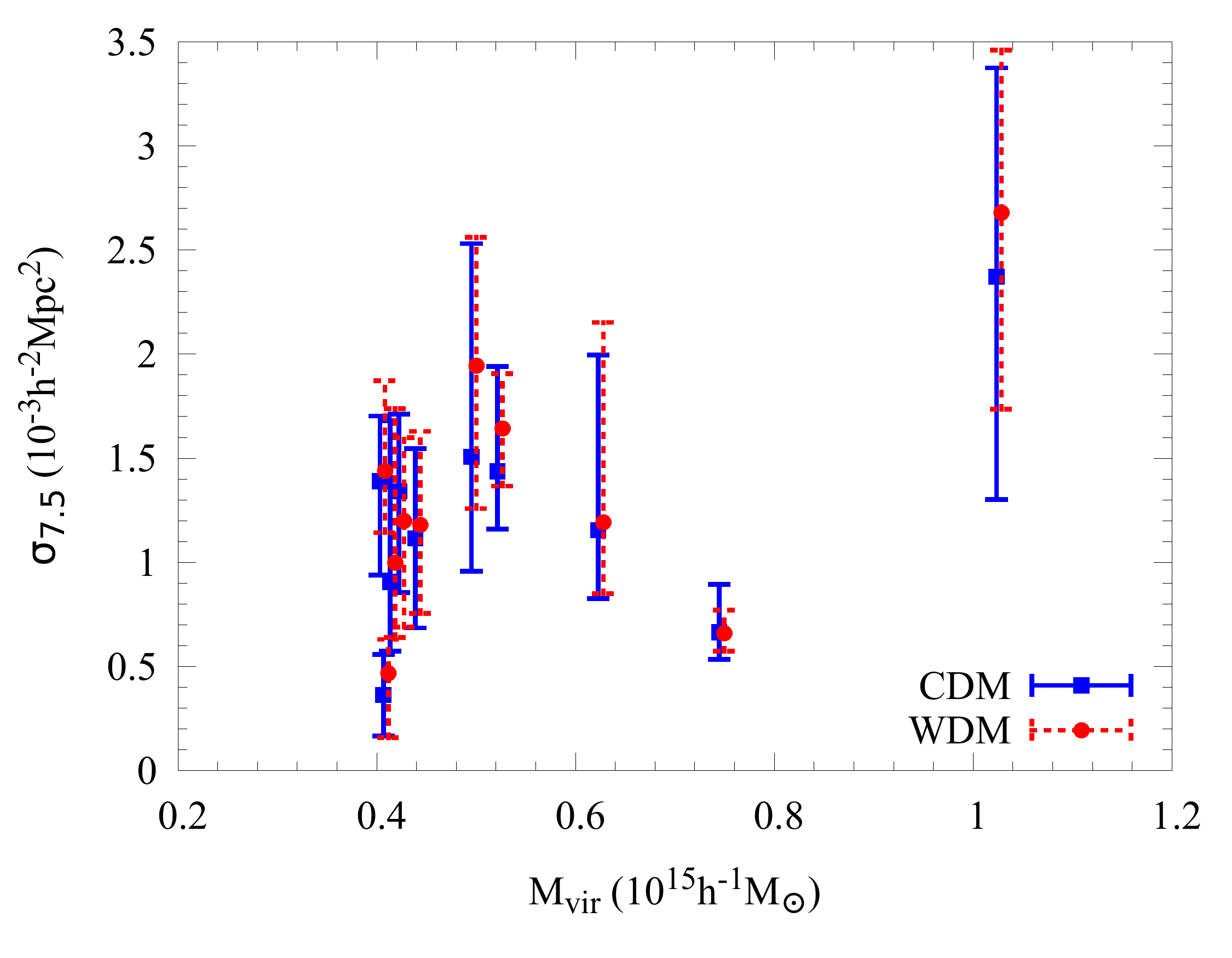}
  \caption{The median cross section for giant arcs along with the 16th and 84th percentile over the 150 los as a function of the virial mass of CDM version of clusters (blue squares) and WDM counterparts (red circles).}
\label{fig:sigma_m} 
\end{figure}

\begin{figure}  
  \centering  
  \includegraphics[width=0.99\columnwidth]{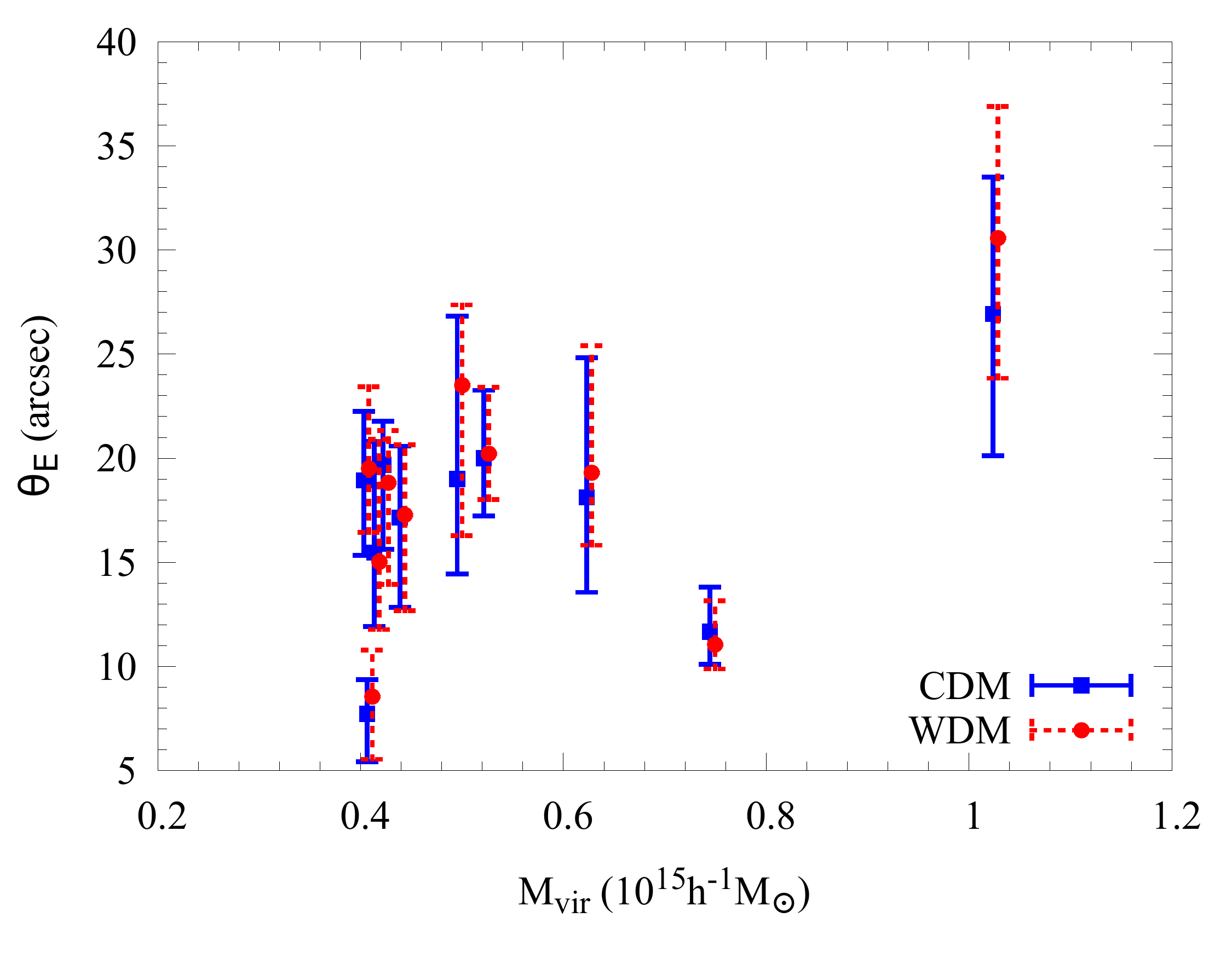}
  \caption{The median Einstein radius for giant arcs along with the 16th and 84th percentile over the 150 los as a function of the virial mass of CDM version of clusters (blue squares) and WDM counterparts (red circles).}
\label{fig:theta_m} 
\end{figure}

\subsection{Lensing Cross Sections \& Einstein Radii}
The length-to-width ratio $L/W$ of the corresponding images is measured by means of the eigenvalues of the magnification tensor $\lambda_r$ and $\lambda_t$. Two types of images are expected: images elongated in the radial direction of critical line which occur when $|\lambda_t|$ is larger than $|\lambda_r|$ and images elongated in the tangential direction with respect to critical line that happen when $|\lambda_r|$ is larger than $|\lambda_t|$.
So, any image with $\left|\frac{\lambda_t}{\lambda_r}\right|$ or $\left|\frac{\lambda_t}{\lambda_r}\right|$ greater than some physically meaningful threshold $\eta$, would be classified as a tangential or radial arc respectively. This technique for computing the lensing cross section was firstly proposed by \cite{2006A&A...447..419F}. The elongation threshold throughout this work is chosen to be $\eta=7.5$ as previously considered in the literature (e.g. \citealt{2005A&A...442..405P,2012A&A...547A..66R}). Higher elongation thresholds lead to a smaller number of giant arcs and less elongation thresholds are very sensitive to the ellipticity of sources \citep{2012MNRAS.427..533K}. 

\par 
Figure \ref{fig:sigma_PDF} shows the probability density function of cross sections of both CDM and WDM samples. The upper panel shows the probability of cross section due to all arcs while the middle and lower panels show the probabilities of cross section due to tangential and radial arcs, respectively. The error bars represent the statistical error for each bin and the sub-plots show the difference between the two distributions. This figure illustrates that the WDM clusters are slightly stronger lenses than the CDM counterparts even though they are less concentrated and contain fewer substructures. The figure also illustrates that the WDM clusters have a more significant tail of large cross sections for radial arcs, which means that the radial arcs are more common in the WDM clusters than in the CDM counterparts.

\par
This difference is also seen in the distribution of Einstein radii, as shown in figure~\ref{fig:theta_PDF}. We see that the WDM clusters produce slightly larger Einstein radii than their CDM analogues.

\par 
As we use the same orientations with respect to the los for both cosmologies, we plotted the lensing cross sections and Einstein radii of the WDM version as a function of those two lensing quantities for the CDM counterpart, as shown in Figure \ref{fig:coldvswarm_sigma} and \ref{fig:coldvswarm_theta}. Again, these two figures illustrate that the WDM version of our sample of clusters produces higher lensing efficiency than their CDM analogues for most of the 1500 los as they lie above the unity line. We found that 64 $\%$ and 60 $\%$ of the lines of sight produce higher lensing cross sections and Einstein radii, respectively, in the WDM version of clusters.    

\par
In Figure~\ref{fig:fitting}, we show the tight correlation between the lensing cross sections and Einstein radii for CDM and WDM clusters. The color bar of the distributions represents the mean shear on the grid for each los. The figure illustrates that the WDM clusters can produce stronger shear field than the CDM counterparts even they have fewer substructures. This figure also illustrates that the lensing cross section and Einstein radius are strongly dependent on the lensing shear, we also found a similar behaviour when the distributions are coloured with the mean convergence on the grid. For this reason, the lensing properties of simulated clusters must be studied by considering different lines of sight. Each line of sight for a particular cluster has a unique convergence and shear fields due to the triaxiality of haloes and the existence of substructures, hence we get a unique Einstein radius and cross section for individual lines of sight. 
The green squares in the lower panel of Figure \ref{fig:fitting} represent the observed sample of clusters from \citep{2011A&A...530A..17M}. Despite the slight enhancement in the lensing efficiency when considering WDM cosmologies, it appears that none of our simulated clusters can reproduce the observed data of 4 clusters in \cite{2011A&A...530A..17M} which have very high Einstein radius and cross section for giant arcs. 

\par
The median cross section and Einstein radius over the 150 los along with the 16th and 84th percentiles are plotted against the virial mass of clusters as shown in Figures \ref{fig:sigma_m} and \ref{fig:theta_m}. We found that eight of the WDM clusters produce higher lensing cross sections than their CDM counterparts. We also found that seven of the WDM clusters produce larger Einstein radii than the CDM analogues. These figures show that the WDM version of most of the clusters produced significantly higher lensing efficiency in comparison to its CDM counterpart. The figures also demonstrate that the variations of the lensing cross sections and Einstein radii of most clusters are significantly wide due to the triaxility of clusters and existence of big subhaloes. This means that the number of lines of sight used in this paper are fairly enough to study the lensing characteristics of galaxy clusters. We also think that the clusters that have narrow distributions are probably more spherical and have fewer big subhaloes. The WDM version of the cluster c10 which produces a higher lensing cross section, has actually a slightly smaller Einstein radius than its CDM counterpart. This is most likely due to the effect of faraway substructures from a cluster's core, if these subhaloes are massive enough, they can develop their own critical lines and caustics, such that they can contribute to the cross section of the clusters, while their contribution to Einstein radius is negligible. We found that the cross section and Einstein radius of the WDM version of our sample could be boosted by up to $\sim$ 29$\%$ and $\sim$ 24$\%$, receptively, in comparison to its CDM counterpart. 

\par 
We perform a least squares fit using $\log(\sigma_{7.5})=m\log(\theta_E)+b$ to the data of both samples. The slope, intercept and the correlation coefficient of the linear fitting are shown in Figure \ref{fig:fitting}. The black lines in Figure~\ref{fig:fitting} show a linear fit to each data set. We found a shallower slope and higher normalisation in the WDM cosmology in comparison to the CDM one. This suggests that the contribution of the source ellipticity to the lensing cross section is more important for the WDM version of the clusters, which is another evidence for the effect of larger size subhaloes in the WDM clusters. These large substructures significantly modify the shear field such that the effect of sources ellipticity is more significant. 

We compare our fit for the CDM sample with those from \cite{2011A&A...530A..17M} and \cite{2012MNRAS.427..533K} as shown in green and red lines, respectively, in the upper panel of Figure \ref{fig:fitting}. Our fit agrees with that of \cite{2011A&A...530A..17M}, even though we still get a shallower slope and higher normalisation. This can be due to the cluster lens redshift, our clusters are simulated up to redshift $z=0$ but we place them at redshift $z_L=0.3$, whereas the clusters in \cite{2011A&A...530A..17M} were taken from simulation at redshift $z>0.5$. Also, it can be attributed to the adopted technique for computing the lensing cross section, the length-to-width ratio of images in this work is quantified by means of the eigenvalues ratio, whereas the giant arcs in \cite{2011A&A...530A..17M} was identified by fitting different geometrical figures to the images, they considered ellipses, circles, rectangles and rings, and then they pick up the figure whose circumference matches best with that of the image under consideration. A comparison between the eigenvalues technique for computing the lensing cross section and that adopted in \cite{2011A&A...530A..17M} has been done by \cite{2012A&A...547A..66R}. These authors have found that the geometrical figures fitting method in \cite{2011A&A...530A..17M} fits $99 \%$ of the lensed arcs to rectangles and that method gives identical results to the eigenvalues method if only ellipses are fitted to the lensed arcs. This means that the geometrical fitting method gives a higher lensing cross section than the eigenvalues method by a factor of $\frac{4}{\pi}$.  The huge disagreement between our relation and that from \cite{2012MNRAS.427..533K} by considering their subsample of relaxed clusters at redshift $z_L=0.25$ can most likely be due to some physical differences between our clusters and theirs (e.g. different abundance of substructures). 
\begin{figure}
    \centering
    \includegraphics[width=0.99\columnwidth]{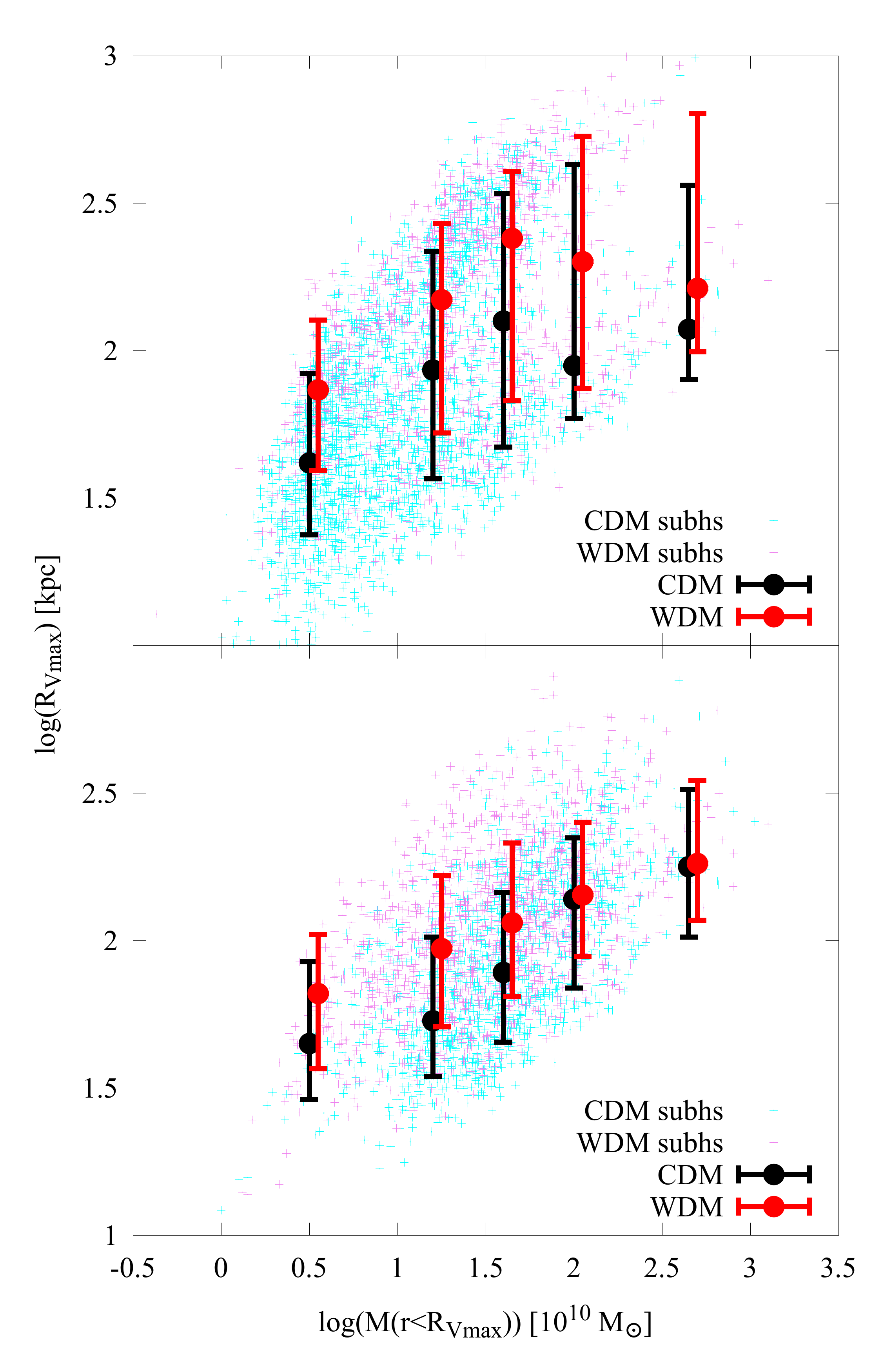}
    \caption{Maximum circular velocity radius as a function of enclosed mass for the subhaloes associated with the most massive haloes in the all ten clusters (upper panel) and for the CDM subhaloes that have matches in the WDM version (lower panel). We show the median and the 16th and 84th percentiles for several mass bins of the CDM distribution (black error bars) and the WDM distribution (red error bars).}
    \label{fig:M_comp}  
\end{figure}

\begin{figure}
    \centering
    \includegraphics[width=0.99\columnwidth]{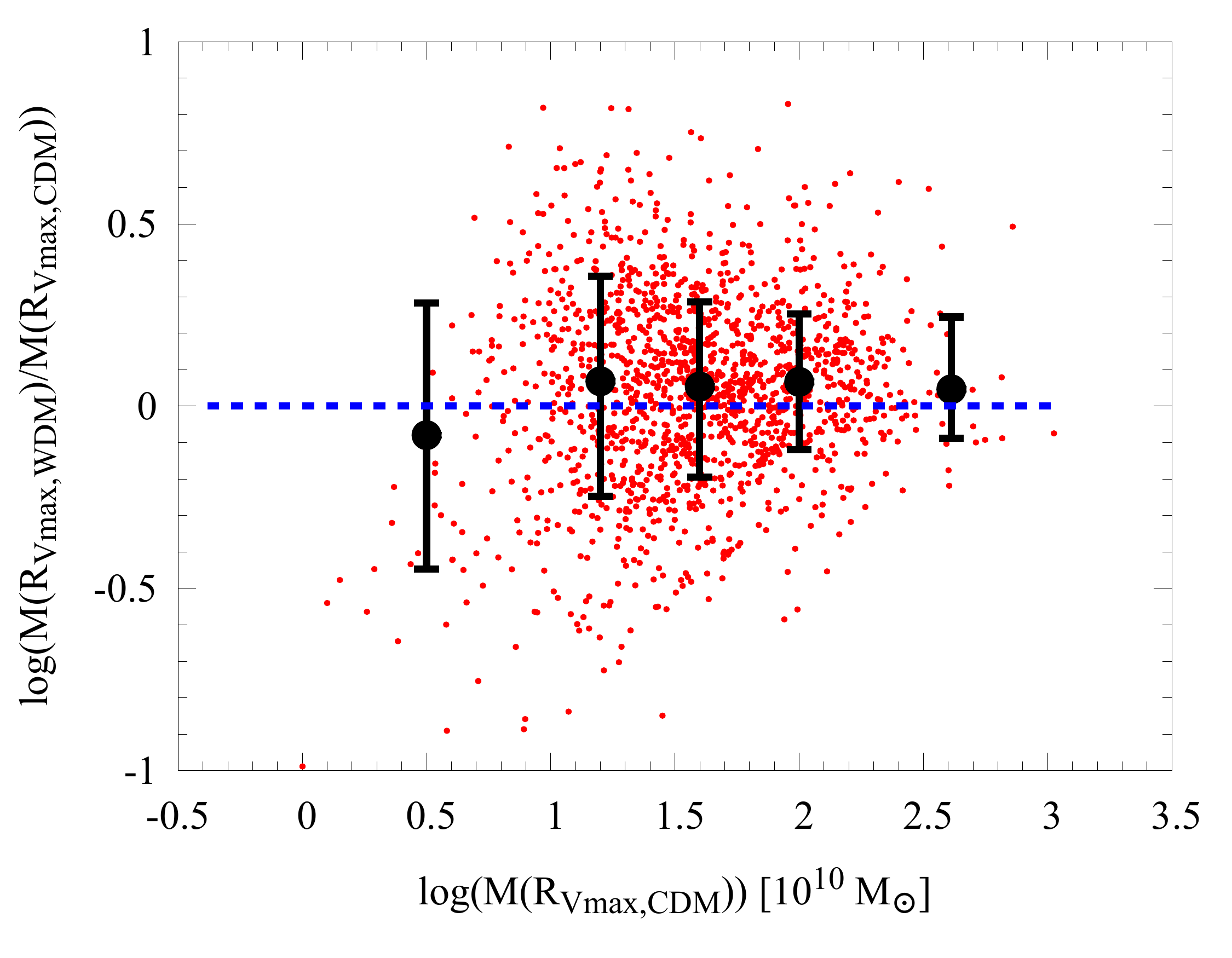}
    \caption{Ratio between WDM and CDM subhalo counterparts of the mass enclosed at the maximum circular velocity radius along with median, 16th and 84th percentiles of several mass bins.}
    \label{fig:M_ratio}  
\end{figure}

\begin{figure}
    \centering
    \includegraphics[width=0.99\columnwidth]{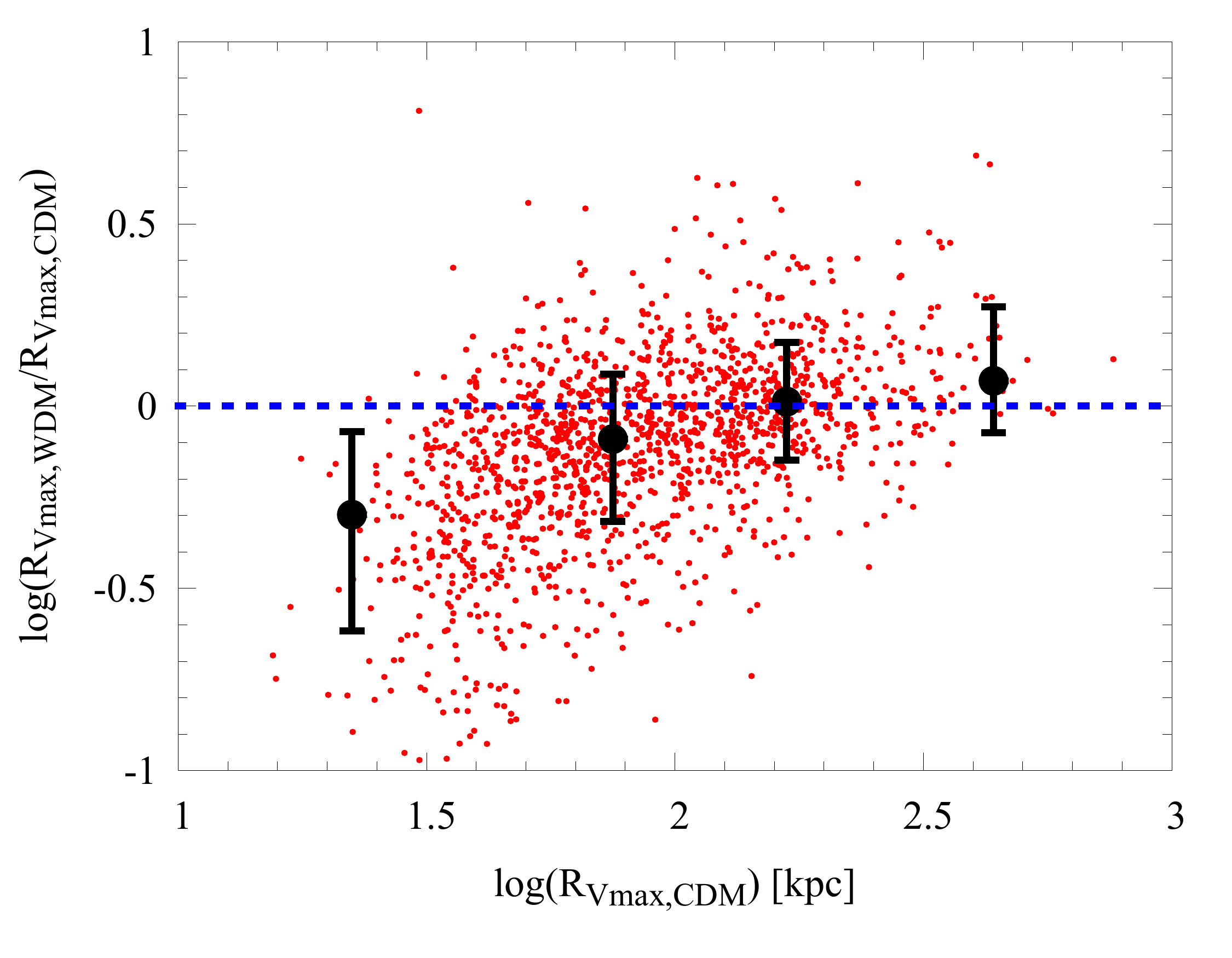}
    \caption{Ratio between WDM and CDM subhalo counterparts of the maximum circular velocity radius along with median, 16th and 84th percentiles of several radial bins.}
    \label{fig:R_ratio}  
\end{figure}

\begin{figure}
  \centering  
  \includegraphics[width=0.99\columnwidth]{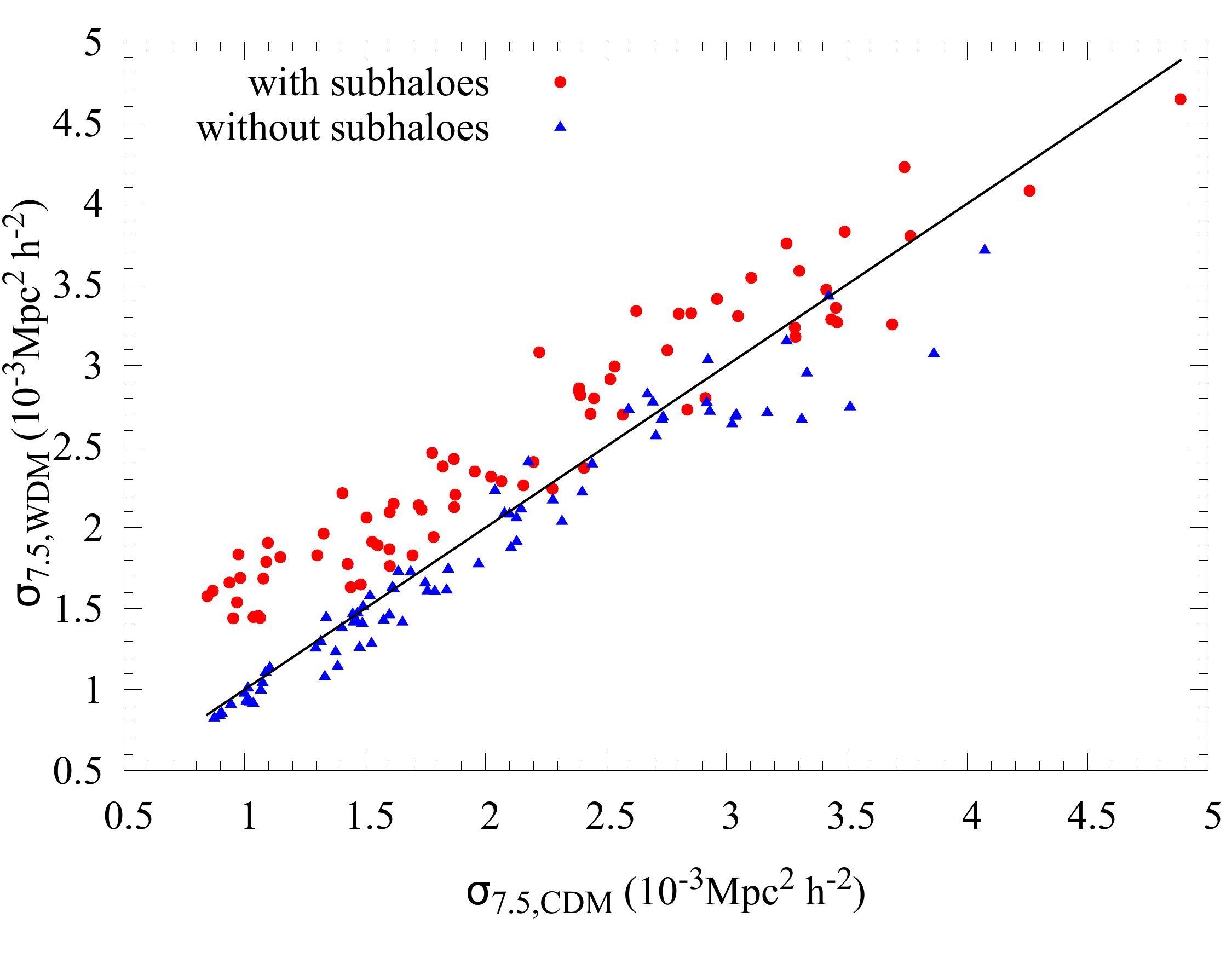}
  \caption{The lensing cross section of the WDM version of the most massive cluster (i.e. c1) as a function of the cross section of the CDM counterpart with and without subhaloes, red circles and blue triangles, respectively. The plot is produced by considering the same orientations with respect to the los. The black solid line is the unity line.}
\label{fig:no_subhs}  
\end{figure}

\section{WDM (slightly) stronger lenses than CDM?}
This result might seem initially counter intuitive since WDM clusters contain less substructures and CDM clusters are more concentrated. A more centrally concentrated mass distribution will produce stronger convergence field, and more substructures will increase the shear field. Based on this naive prediction, the CDM clusters should produce larger cross sections and Einstein radii than the WDM counterparts. However, our results show that eight of the WDM clusters produce higher lensing efficiency than the CDM version of the clusters. We found that the key difference in the WDM clusters is the existence of more massive and more extended substructures than those in the CDM counterparts. Those subhaloes significantly enhance the shear field of the host cluster if they orthogonally aligned with the line of sight and they increase the convergence if they aligned along the line of sight.      
We also found that these WDM subhaloes significantly perturb the radial critical line such that the lensing cross section for radial arcs is higher in the WDM  clusters than the CDM ones.  

To pin down the effect substructures have on the lensing distribution, we examine their internal properties and mass growth. In the companion study \cite{Elahi2014}, we identify all bound (sub)haloes of every snapshot of our simulated clusters using {\sc VELOCIraptor} \citep{2011MNRAS.418..320E}. This code identifies haloes using a 3D Friends-of-Friends (FoF) algorithm with a linking length of 0.2 times the interparticle spacing, then identifies dynamically distinct phase-space structures residing within each halo. We briefly present here some of the salient findings concerning substructure here (for more details see the companion paper \citealp{Elahi2014}). The upper panel of Figure \ref{fig:M_comp} shows the maximum circular velocity radius as a function of enclosed mass for the subhaloes associated with the most massive halo in the 10 clusters of CDM (cyan crosses) and the WDM (violet crosses) versions. The lower panel on the other hand shows the same plot for the CDM subhaloes that have analogues in the WDM version of clusters. The cross catalogue is made by identifying each (sub)halo $i$ in catalogue A that shares particles with a (sub)halo $j$ in catalogue B and the merit of the initial matches is determined from $\mathcal{M}_{ij}=N^2_{A \cap B}/(N_A N_B)$ using the halo merger tree code of {\sc VELOCIraptor}, which is a particle correlator (see \citealp{2013MNRAS.436..150S} for more details of this code). The figure illustrates that on average for a given mass, WDM subhaloes are more likely to have larger sizes than the CDM analogues. Figure \ref{fig:M_ratio} shows the ratio between WDM and CDM subhalo counterparts of the mass enclosed at the maximum circular velocity radius. The error bars represent the median, 16th and 84th percentiles of several mass bins. The figure illustrates that the median of the logarithmic ratio is larger than zero, which means the WDM subhaloes are more massive than those in the CDM clusters. Figure \ref{fig:R_ratio} shows the ratio between WDM and CDM subhalo counterparts of the maximum circular velocity radius. The error bars represent the median, 16th and 84th percentiles of several radial bins. Again, the figure demonstrates the possibility of having more extended WDM subhaloes than the CDM counterparts.

\par 
In order to check whether the boost in the lensing efficiency of the WDM clusters comes from the subhaloes, we compute the lensing properties of both versions of the most massive cluster (i.e. c1) with and without projecting the particles in the subhaloes on the lens plane. Figure \ref{fig:no_subhs} shows the cross section of 75 los of the WDM version versus the cross section of the CDM counterpart by considering subhaloes (red circles) and excluding them (blue triangles). The figure demonstrates that by including the substructures, the WDM version produces higher lensing efficiency for 84 $\%$ of the lines of sight. This fraction drops off to 25$\%$ when excluding the particles associated with substructures. This clearly proves that the higher lensing efficiency of the WDM clusters is due to the contribution of subhaloes.  

\section{Discussion and Conclusions}
\label{sec:discus}
The lensing properties for two samples of simulated clusters in the WDM and CDM cosmologies have been studied. Based on the characteristics of WDM clusters, one would expect them to be weaker lenses than the CDM counterparts as they are less centrally concentrated and have fewer satellites than CDM analogues. The results of this study show that the WDM clusters are slightly more efficient lenses at producing giant arcs than the CDM counterparts.

We found that the key difference in the WDM clusters that significantly enhances their lensing efficiency is the existence of more extended and more massive substructures than those in the CDM clusters. Those more massive substructures significantly enhance the shear and convergence fields which in turn lead to a higher lensing efficiency. 

However, despite the enhancement in the lensing efficiency, WDM alone cannot account for the differences seen between theory and observation. 
 
A more robust comparison regarding the effect of small-scale substructures can be done by adopting recursive-TCM projection algorithm that has recently been proposed by \cite{2013arXiv1309.1161A}. The advantage of this method is that there is no need for Gaussian smoothing, and therefore small-scale substructures that are washed away by smoothing remain. However, the contribution of such small-scale substructures to the lensing efficiency is probably not significant.

We found that eight of the WDM clusters produce higher cross sections and Einstein radii than their CDM counterparts. However, we found that the WDM version of the cluster c10 produces higher cross section but smaller Einstein radius than its CDM counterpart. Similarly, the WDM version of the cluster c2 produces smaller cross section but larger Einstein radius than its CDM counterpart. This can be attributed to the existence of some substructures which are strong enough and relatively faraway from the center of a cluster to develop their own critical lines and caustics. In such a case, these substructures contribute to the lensing cross section but not for Einstein radius as we compute Einstein radius only for the core of clusters.

We have fit the function $\log(\sigma_{7.5})=m\log(\theta_E)+b$ to our data and found that the relation for the WDM sample has a shallower slope and higher normalisation than the CDM counterpart. This is again due to the effect of substructures which substantially affect the shear field. The source ellipticity becomes more significant for the lensing cross section  of the WDM lenses due to their larger subhaloes.  
By comparing our results to those of \cite{2011A&A...530A..17M} and \cite{2012MNRAS.427..533K}, we find a good agreement between our fitting and that from \cite{2011A&A...530A..17M}. However, our fit disagrees with that from \cite{2012MNRAS.427..533K}, this is can be due to some physical differences of our clusters and theirs. 

We conclude that despite the slight enhancement in the lensing efficiency of WDM clusters, they fail to explain the arc statistics problem as none of our clusters can reproduce the observed cross sections and Einstein radii of four clusters in \cite{2011A&A...530A..17M}. As a possible solution, we are going to study the effects of baryonic physics on the lensing properties of WDM clusters. Furthermore, we are planning to analyse the lensing characteristics of clusters of galaxies drawn from other cosmologies, such as coupled dark matter-dark energy models.               
         
\section*{Acknowledgements}
HSM is supported by the University of Sydney International Scholarship and would also like to thank Matthias Redlich for the helpful discussions at the very early stages of this project. PJE is supported by the SSimPL programme and the Sydney Institute for Astronomy (SIfA), DP130100117 and DP140100198. CP is supported by DP130100117, DP140100198, and FT130100041. MK acknowledges support by the DFG project DO 1310/4-1.

\bibliographystyle{mn2e}
\bibliography{paper.bbl}

\label{lastpage}
\end{document}